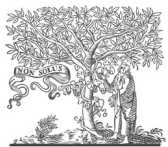
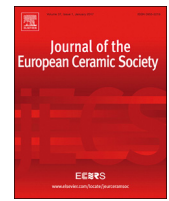

Short communication

# 3D crystallographic alignment of alumina ceramics by application of low magnetic fields

Alexander S. Sokolov*, Vincent G. Harris

*Department of Electrical and Computer Engineering, Northeastern University, Boston, MA 02115, USA*

ABSTRACT

Non-cubic crystals exhibit anisotropic physical and functional properties. Microscopic crystallites as constituents of polycrystalline materials are randomly oriented, thus polycrystalline ceramics lack the anisotropic properties of their monocrystalline counterparts. We propose a technology that exploits the synergy between magnetic alignment and colloidal ceramics processing, and enables to independently tune the orientation of grains in different sample regions by infinitesimal magnetic fields (< 10 mT). The grain pivot mechanism enables the emulation of single crystals, as well as the creation of large complex-shaped ceramic elements with designed crystallographic landscapes and spatially and directionally tuned properties. Ultra-high magnetic response arises from magnetic shape anisotropy of platelet-shaped seed crystallites coated with small amounts of iron oxide nanoparticles. To control crystallographic growth directions during subsequent annealing procedures, the seeds are dispersed and aligned in a matrix of chemically identical, but much finer spherical particles. This technology opens an avenue to remarkably improve structural and functional properties of ceramic elements employed in numerous industrial applications.

## 1. Introduction

### 1.1. Background

Ceramics constitute a broad class of materials defined generally as inorganic non-metallic solids whose ions are held together predominantly by covalent and ionic bonds. The degree of crystallinity may vary from highly ordered to amorphous. Functional engineering ceramics mostly have a crystalline structure, for example zirconium dioxide, silicon carbide, or aluminum oxide. Crystals consist of atoms forming near-perfect periodic arrangements. The microscopic atomic structure is often described in terms of its unit cell – an elementary building block of atoms that when repeated through translational and rotational symmetry create a bulk crystal. Unit cells determine a crystal's symmetry where the numerous symmetries possible in the 3D space are formally divided into 14 Braivais lattices.

Cubic systems are isotropic and possess the highest of crystal symmetries. Hexagonal, trigonal, rhombohedral, tetragonal, and monoclinic systems have anisotropic crystallography with preferred – or principal – axes that are intrinsically of higher symmetry over other axes. Many physical properties of such non-cubic crystals can often be anisotropic [1] – i.e., depend on crystallographic orientation; for example, thermal, optical and electrical conductivity, as well as mechanical properties such as strength, hardness, and toughness. This degree of anisotropy is crucial for many applications. Hexagonal boron nitride [2] crystals depict a striking illustration. Their thermal conductivity along the C (long axis) and A (short axis) planes differ by a factor of 20, the higher value in C plane exceeds the one of stainless steel by a factor of 30. This is due to the different natures of interatomic bonding in these two crystallographic directions – covalent in C plane and Van der Waals along the C axis respectively.

Macroscopic non-cubic single crystals possess unique anisotropic properties that find countless applications. However, it is difficult, expensive, and often impossible to produce such crystals as large or complex shaped. On quite the contrary, most polycrystalline ceramics are compliant with a variety of near-net-shape processing techniques, including slip- and gelcasting, spray forming, and injection molding. Nevertheless, microscopic crystallites as constituents of such polycrystalline materials are usually randomly oriented, thus polycrystalline ceramics lack the anisotropic properties of their monocrystalline counterparts.

It has been a longstanding goal of the materials science community to establish control over the spatial orientation of crystallites in polycrystalline ceramic materials. Not only does it allow one to emulate single crystals, but also to produce ceramic parts with locally varying crystallographic orientations and therefore properties, for example to achieve reinforcement in mechanically loaded directions. This technology is of benefit to numerous industries and creates a path to realizing remarkable improvement to structural and functional attributes of ceramic materials used in biomedical implants, turbine blades, transparent ceramics, antiballistic armament, photoluminescent materials, semiconductors and solid-state electronic substrates, among others.

* Corresponding author.
*E-mail address:* sokolov.al@husky.neu.edu (A.S. Sokolov).

*1.2. Conventional magnetic alignment*

Anisotropic atomic structure in non-cubic crystals often results in magnetocrystalline anisotropy. When placed in an external magnetic field, single-domain crystallites with uniaxial or triaxial magnetocrystalline anisotropy experience torque that rotate these particulates aligning their easy magnetization axes along the direction of applied magnetic field. Favorable directions of spontaneous magnetization – i.e., easy axes – coincide with certain, usually principle, crystallographic axes, that allows the control of the spatial crystallographic orientation of non-cubic crystallites.

This approach is routinely used to produce "magnetically hard" ceramic (i.e., hexaferrite) magnets [3–5]. Large and anisotropic magnetic susceptibility of hexaferrite particles results in significant torque and allows for the efficient alignment of grains by weak magnetic fields applied during conventional powder metallurgy processes. In the case of hexaferrites, micron-sized, single-crystal, single-domain ferrite particulates are suspended in a solution, then aligned by a magnetic field of the order of 1 T, compacted in a desired shape, and sintered to form a solid ceramic magnet with a high degree of crystalline orientation that makes it difficult to demagnetize.

It is desirable to similarly control the crystalline orientation in engineering polycrystalline ceramics by a magnetic field. Non-magnetic, non-cubic ceramics are in fact para- or diamagnets, and also possess magnetocrystalline anisotropy. Its absolute value, however, is significantly lower than that of popular ferro- or ferrimagnetic materials, therefore successful alignment of such crystallites requires remarkably high magnetic fields, often greater than 10 T.

For instance, the rhombohedral structure of diamagnetic alpha-alumina results in a magnetic anisotropy $\Delta\chi$ of the order of $1 \times 10^{-9}$ emu/g [6]. Such a small anisotropy could be considered negligible. However, the spreading of superconducting magnets capable of magnetic fields in excess of 10 T makes it feasible to align such non-magnetic ceramics similarly to "magnetically hard" ceramic magnets. The energy of the magnetic anisotropy for submicron spherical alumina particles in a 10 T field exceeds the energy of thermal motion at room temperature by about an order of magnitude. This approach allows relatively high degrees of crystalline orientation [7–10], but it is not truly practical because a 10 T field can only be applied over small volumes, thus severely limiting sample sizes.

*1.3. Ultrahigh magnetic response of surface-magnetized platelets and whiskers*

Magnetocrystalline anisotropy is not the only viable origin of magnetic anisotropy. A perfect single crystal is formed by a set of discrete translations of its unit cell. Consequently, the crystal and its cell possess discrete translational symmetry and represent equivalent Bravais lattices. As a result, micro- and macroscopic ceramic crystallites may often retain the shape and axes of symmetry of their corresponding unit cells. Such crystallites frequently exist with high aspect ratios; common examples include platelets and whiskers. Shape anisotropy likewise gives rise to magnetic anisotropy. In this case, the longest dimension of a magnetized body, which now coincides with one crystallographic axis, becomes analogous to an "easy" magnetization axis.

It has been shown that nonmagnetic crystallites, including aluminum oxide [11], hexagonal boron nitride [12], and even carbon nanotubes [13], can be coated with iron oxide nanoparticles [14]. Less than 1 vol. % of these magnetic nanoparticles attached to the surface of host crystallites dramatically increases the overall magnetic susceptibility, and, when coupled with a high aspect ratio of the host, allow for magnetic alignment of nonmagnetic crystallites based on their shape anisotropy.

The spatial orientation of high aspect ratio ($\geq 10$) crystallites suspended in a solution and exposed to a dc magnetic field is primarily determined by 3 competing energies – thermal, gravitational, and magnetic. Both gravitational and magnetic energies depend on the angle between the crystallite longest dimension and the corresponding field, and reach minima when said axis becomes parallel to respective fields. A necessary condition for the magnetic alignment is the preeminence of the magnetic energy. Erb et al [11] evaluated and plotted all three energies as functions of the length for rods and diameter for platelets, and discovered that an ultrahigh magnetic response (UHMR) is possible for crystallites with the longest dimension of about 10 um. These authors further determined that, for a given crystallite geometry, the theoretical minimum magnetic field necessary for the alignment declines nearly linearly with the volume fraction of magnetic nanoparticles. UHMR platelets with 0.5 vol. % coating can hypothetically be aligned by a 1 m T field - an intensity that exceeds the Earth's magnetic field [15] by only one order of magnitude; if the coating is shrunk to 0.05 vol. %, the minimum necessary field increases to 10 m T – the intensity of a ferrite-based refrigerator magnet. Subsequent experiments demonstrated good agreement with this theory.

*1.4. Templated grain growth*

Apparently, crystallites of many non-cubic functional ceramics can be synthesized to near optimal UHMR specifications, i.e., aspect ratio $\geq 10$, and the long axis dimension between 5 and 15 um; for example, α-silicon carbide (hexagonal structure) platelets, [16–20] boron carbide (rhombohedral structure) platelets and whiskers [21], boron nitride (hexagonal structure) platelets [22,23], α-aluminum oxide (hexagonal structure) platelets [24–26], zinc oxide platelets or rods [27], and β-silicon nitride (hexagonal structure) platelets [28] have been demonstrated.

These findings potentially allow one to manipulate crystallographic orientations in polycrystalline ceramics by application of surprisingly low magnetic fields. Platelets and whiskers, however, tend to form skeletons and vigorously resist densification. Hence, they cannot be directly used to produce dense ceramic parts via conventional powder metallurgy process. Up until now the described alignment method has only been used to control the orientation of ceramic reinforcement elements in polymer composites [29–36].

We put forward here a methodology that allows to synthesize high density functional ceramics with micron sized grains and crystallographic orientations locally tuned in three dimensions by low magnetic fields.

A small relative amount of high aspect ratio UHMR crystallites can be dispersed in a matrix of chemically identical, but much finer spherical particles, and aligned by applying low magnetic fields throughout the green body formation process via slip- or gel-casting. Subsequently, these larger oriented crystallites will serve as the centers of nucleation for finer particulates during sintering, ultimately resulting in dense ceramics with tailored crystallographic orientations. Low magnetic fields, required to align UHMR crystallites, allow the use of multipolar magnets. Therefore, the ability to create complex spatial and crystallographic directions, in both small and large ceramic elements, is now realizable.

To put this in perspective, anisotropic diamagnetic particles can be directly aligned by magnetic fields in excess of 10 T that can only be induced by superconducting magnets; ferrimagnetic particles exhibit much larger magnetocrystalline anisotropies and get aligned routinely by 1–2 T fields that are induced by large electromagnets; high aspect ratio UHMR crystallites can be oriented by a refrigerator magnet that only produces a few millitesla.

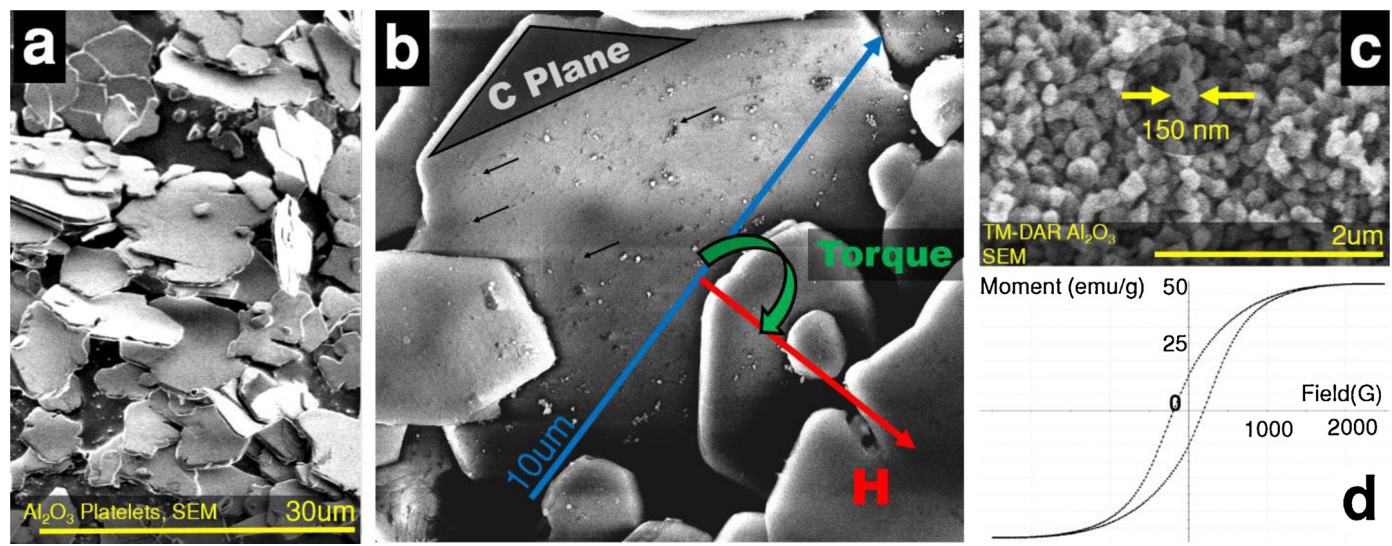

**Fig. 1.** (a) Alusion™ alumina platelets powder, SEM. (b) alumina platelet coated with iron oxide nanoparticles (marked with black arrows). (c) TM-DAR alumina powder, SEM. (d) Maghemite iron oxide nanopowder, M vs H loop.

## 2. Experimental procedure

Arguably the most versatile functional ceramic that finds applications in countless industries – aluminum oxide – has been selected for experimental corroboration of the proposed approach. The thermodynamically stable phase of aluminum oxide – alpha alumina – belongs to the hexagonal crystallographic family, hence its properties exhibit anisotropic behavior. This crystalline form of alpha alumina is also called corundum and commonly occurs in nature. Crystals often have elongated shapes due to its hexagonal close-packed atomic structure, and can be striated on faces. Hexagonally shaped thin plates occur naturally and can also be produced synthetically (Fig. 1).

### 2.1. Preparation of $Fe_2O_3$-coated crystallites

A sample of Alusion™ synthetic single crystalline alumina platelets powder employed in this study has been supplied by Antaria Limited, Australia. Scanning electron microscope (SEM) analysis (Fig. 2a) revealed the platelets to have a mean diameter and thickness of 10 um and 0.2 um, respectively – near perfect UHMR geometries. The crystallographic C-axis is normal to the plane of the platelets. The platelets have been coated with 0.1–1 wt % of maghemite iron oxide nanoparticles according to the following protocols.

Iron oxide nanoparticles and the platelets (host crystallites) were suspended in an alkaline pH aqueous buffer solution (10–20 wt% ceramic content), accumulated opposite surface charges, and became attracted by electrostatic forces. Permanent bonding was formed by van der Waals interactions. Arguably, striated faces of the host crystallites allowed for stronger bonding.

We first introduced the desired amounts (0.1, 0.5, and 1 wt % with respect to alumina platelets) of iron oxide nanoparticles ($\gamma$-$Fe_2O_3$, < 50 nm, Sigma-Aldrich) to a series of Thermo Scientific™ buffers with pH 7, 9, and 10, and sonicated these solutions in appropriately sized beakers for 30 min. to reduce agglomeration.

Then the alumina platelets were introduced and solutions mixed for 30 min. to allow for complete absorption. Finally, solutions were dried, and coated platelets were washed three times in DI water. The most durable coatings were achieved in pH 9 buffers and shown to withstand sonication, strong magnetic fields (~1 T), and dispersants.

Alumina platelets and iron oxide nanoparticles, respectively, acquire positive and negative surface charges in alkaline solutions. This is not always the case – other host crystallites may exhibit negative surface charging. Then iron oxide nanoparticles, preliminary treated with cationic surfactants (for example EMG-605, Ferrotec, Germany [11]), were found necessary to exploit the electrostatic attraction.

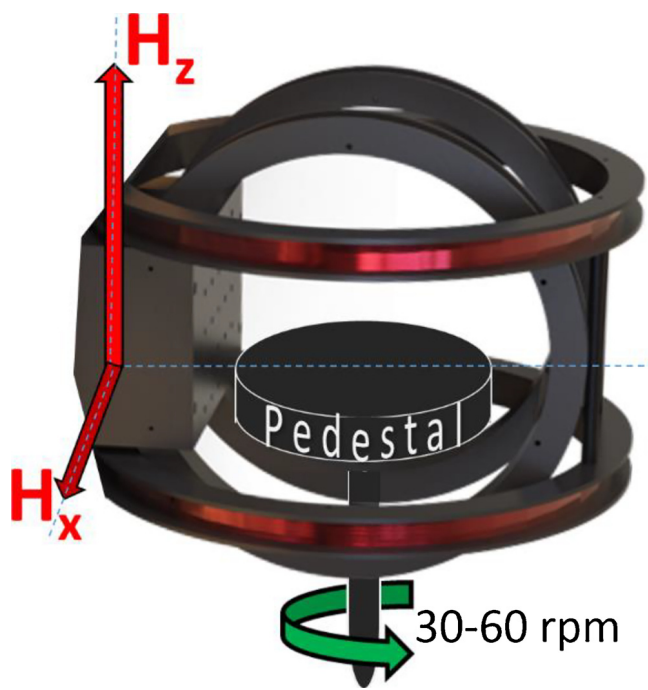

**Fig. 2.** 2-Axis Helmholtz Coils and a rotating sample pedestal. Schematic diagram.

### 2.2. Characterization of $Fe_2O_3$-coated platelets

A SEM micrograph of an alumina platelet coated with 0.1 wt % of iron oxide (Fig. 2b) shows a uniform distribution of nanoparticles on its surface. The starting $\gamma$-$Fe_2O_3$ powder and alumina platelets coated with this powder produced similar hysteresis loops and specific (with respect to iron oxide) magnetizations (Fig. 2d), confirming that no additional effects were involved. Neither showed a truly superparamagnetic behavior.

The response of coated alumina platelets suspended in solution to an applied magnetic field stimuli was directly observed and filmed using an optical microscope in both transmission and reflection modes (Supplemental video 1). Free-floating platelets indeed exhibited the UHMR and rapidly aligned their long axes parallel to the applied magnetic field. Smaller (∼1 μm) particulates present in the solution experienced intense Brownian motion, confirming that this method is not viable for aligning sub-micron crystallites and producing ceramics with a very fine microstructure. It was also noted that the UHMR can be plagued by a phenomenon that is hard to account for within the energy model discussed - the platelets tended to agglomerate, and also stick to the surface of the glass slide. We subsequently employed successfully an anionic ammonium polymethacrylate dispersing agent (DARVAN® C–N, Vanderbilt Minerals, LLC) [37–39] and a vibrating table to compensate for this effect.

### 2.3. Gel-casting in small amplitude magnetic fields

An aqueous gel-casting system based on a water soluble copolymer of isobutylene and maleic anhydride has been used to produce complex shaped green bodies. The slurries containing desired amounts of the UHMR platelets and alumina matrix powder (TM-DAR α-Al2O3, 100 nm, 99.99%, Taimicron, Japan) [40], and, sometimes, a combination of sintering additives (TiO$_2$ + CuO), were poured into molds.

A sample of the polymer, marketed as ISOBAM 100 series, has been provided by Kuraray America Inc. A detailed description of the system can be found elsewhere [41]. The polymer allows for low viscosity aqueous ceramic slurries with high solids loading and therefore befits the magnetic alignment.

The slurries were prepared by first mixing a desired amount of distilled water and a dispersing agent Darvan C–N (0.014 cm$^3$ per gram of alumina regardless of the solids loading). Then the UHMR platelets and alumina matrix powder (TM-DAR α-Al$_2$O$_3$, 100 nm, 99.99%, Taimicron, Japan) (Fig. 2c) were introduced and ball-milled for one hour. 0.2 wt% (with respect to alumina) of ISOBAM-110 (molecular weight 160,000–170,000) was separately dissolved in a small amount of distilled water, introduced to the slurry, and then followed by one hour of ball milling. Prepared slurries were outgassed in a vacuum chamber, poured into molds, and outgassed again. The total solids loading varied between 60–68%.

The molds were placed on a vibrating pedestal and exposed to approximately 10 mT magnetic fields of desired field configurations – induced by either electromagnets or permanent ferrite magnets, or both (Fig. 2). The vibration was turned off after 1 min. Supplemental video 1 shows that some platelets may stick to the surface of the mold. The vibration was intended to suspend these platelets. It also helped to temporarily slow down the gelation process and allowed for a more efficient alignment. The slurries then gelled in said magnetic fields with virtually no shrinkage, permanently freezing in the spatial orientation and disposition of the platelets. The initial gelation time of appropriately prepared slurries was less than 30 min. (Reduced solid loadings in the slurries can increase the gelation time dramatically [41–43]). The samples fully hardened in 12 h, and were subsequently removed from the molds and kept in a constant temperature furnace (Yamato DKN402C) at 70 °C for 48 h to remove the residual water.

Helmholtz Coil electromagnets (Fig. 2), such as GMW Model 5451, are perfectly suited for the alignment of the UHMR platelets. Such magnets provide large volumes of high uniformity and the maximum magnetic fields of about 50 mT.

Control over the orientation of seed platelets within the matrix subsequently allows the control of crystallographic orientation of crystallites composing the post-fired ceramics. The grain orientation distributions are restrained by the magnetic field distributions that are induced within the mold. Complex architectures, enabled by multipolar magnets, consist of two types of basic building blocks with distinctive orientations. These orientations can be described as "random in plane" (Fig. 4b) and "out of plane" (Fig. 4a).

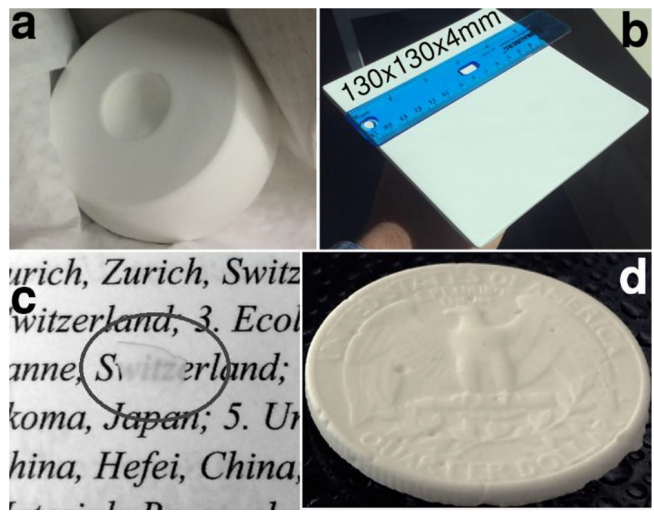

Fig. 3. Gel-casted ceramic elements. (a) 40 mm alumina sphere with a milled face and a 12 mm hole (geometry similar to ceramic femoral heads used for total hip arthroplasty). (b) Alumina tile. (c) 0.5 mm thick translucent alumina specimen (out of plane C axis). (d) Alumina coin replica (U.S. Quarter).

A locally linear magnetic field aligns the platelets along their longest axes, which are normal to their crystallographic C-axes. The platelets, however, are free to rotate around their longest axes. Thus, C-axes of the platelets are randomly distributed in the plane normal to the applied magnetic field (Fig. 4b).

A rotating magnetic field can be used to obtain control over the second degree of freedom of the platelets. In this case, the C-axes aligns parallel to the axis of rotation of the applied magnetic field (Fig. 4a). In this study, the rotating magnetic fields have been produced by rotating the sample pedestal at 30–60 rpm in linear static magnetic fields induced by Helmholtz Coils (Fig. 2). A more advanced setup should utilize 3-axis Helmholtz Coils and a programmable power supply to allow for rotating magnetic fields in any planes. Then, the pedestal and samples may remain static.

More complex architectures with high spatial resolutions can be obtained by employing platelets coated with varying amounts of magnetic nanoparticles that respond selectively to magnetic fields of certain intensities. [28]

A variety of green bodies were produced, including spheres and tiles (Fig. 3). The green bodies appeared to be easy to machine and strong enough to withstand drilling and milling (Fig. 3a). Optical microscopy was used to verify the uniform spatial distribution and desired orientation of the platelets within unfired ceramic parts.

### 2.4. Sintering and crystallographic texture development

Pure TM-DAR alumina powder reached near theoretical density in 1 h at 1300 °C (< 0.02% residual porosity, as measured by a Micromeritics AccuPyc 1330 2.01 pycnometer.). 5 wt% of UHMR platelets increased the necessary sintering temperature to 1500 °C. When 10 wt% of the platelets was present, a combination of sintering additives (0.45 mol % TiO$_2$ + 0.45 mol % CuO with respect to alumina, Titanium (IV) Oxide, < 25 nm particle size, 99.7% purity, Sigma-Aldrich, and Copper (II) Oxide, < 50 nm particle size, Sigma-Aldrich) was used to maintain the minimum required sintering temperature at 1500 °C.

To determine the optimal sintering parameters, as well as to study the evolution of the crystalline orientation, all types of the green bodies were sintered at 1100, 1200, 1300, 1400, 1500, and 1600 °C. The dwelling time varied between 1 and 6 h, and the ramping rate remained constant at 5 °C /min.

## 3. Results and discussion

Much finer matrix particulates nucleate on platelet-shaped seeds throughout the course of sintering in a process known as templated [28] grain growth (TGG), resulting in dense ceramics with desired spatial distributions of crystallographic orientations of grains. The process largely depends on the fraction of seeds in the green body. This parameter was varied between 1 and 25 wt. % with respect to the mass of the matrix powder. A platelet content between 5–10 wt. % were found to be optimal. Lower concentrations resulted in lower degrees of crystalline orientation, and higher concentrations result in low density compacts after sintering.

The degree and type of crystalline orientations that developed throughout the sintering as a result of the templated grain growth was investigated by XRD, optical, and electron microscopy.

Platelets served as nucleation centers for matrix powder, and both SEM and optical micrographs reveal that the final grains mostly retained anisotropic platelet-like shape and were aligned in the intended directions.

The "out of plane" orientation essentially emulates the single crystal and should produce an X-ray diffraction patterns similar to sapphire. X-ray diffraction peaks at low interplanar angles (e.g., < 006 > , < 1010 > ) are characteristic to the crystallographic C-plane being nearly parallel to the face of the sample under test. Crystallographic planes normal to the C-plane have characteristic X-ray peaks at high interplanar angles (e.g., < 110 > , < 030 > ). A perfect C-plane sapphire wafer only reflects the < 006 > peak. A quasi-sapphire crystalline structure was produced by sintering an 'out of plane' aligned green body at 1500 °C for 6 h. The degree of crystalline orientation was calculated as $P = I_{006}/(I_{006}+I_{110})$ and reached 99%. $I_{006}$ and $I_{110}$ are the relative intensities of < 006 > and < 110 > XRD peaks. Fig. 5 shows the evolution of crystalline orientation with sintering temperature.

There is no simple way to quantify the degree of crystalline orientation for "random in plane" aligned samples. If the C-axes of the grains are in a plane parallel to the face of the sample under test, then the < 006 > and < 1010 > peaks (low interplanar angles) should disappear, and the highest intensity should be reflected at the < hk0 > (high interplanar angles) peaks. Fig. 4h, therefore, confirms a qualitatively high degree of crystalline orientation in a "random in plane" sample.

Pre-fired iron oxide content varied between 0.1-0.005 wt%, and the green bodies had a slight orange color. Sintered ceramic parts, including the ones doped with 0.45 mol % $TiO_2$ + 0.45 mol % CuO, became bright white and translucent, suggesting that iron oxide was no longer present in its original form after sintering. XRD analysis also did not unearth the presence of iron oxide after annealing.

We additionally performed EDX spectroscopy to analyze the impurities before and after sintering. The original composition of the tested samples was the following: 100 wt. % TM-DAR alumina nanopowder + 10 wt. % UHMR alumina platelets + 0.05 wt. % $Fe_2O_3$ + 0.45 mol % CuO + 0.45 mol % $TiO_2$ + 0.2 wt. % ISOBAM-110.

Although the matrix powder was the high purity (99.99%) TM-DAR alumina, the analysis of the pre-sintered samples revealed the presence of glass forming impurities, such as Si, Na, and also Cl and Ti, in weight concentrations of the order of 1% (Fig. 5(e)). These elements are presumed to exist in oxidation states. While Ti was intentionally introduced as a sintering additive, other elements must have originated from the UHMR platelets because their purity, compared to the matrix, was much lower.

EDX analysis of the pre-sintered samples only involved the generation of X-ray spectra from fairly large scan areas of 100 by 100 μm. The background continuum typically does not allow to detect the characteristic X-rays from elements below approximately the 0.1 wt. % level, therefore the presence of iron could not be revealed.

The spectra of said samples changed significantly after annealing at 1500 °C for 6 h. The characteristic peaks of all impurities disappeared. The spectra were collected from the grains, grain boundaries, and large scanning areas, and appeared virtually the same.

Therefore, XRD and EDX spectroscopy do not allow to quantify trace amounts of impurities in the sintered samples. If necessary, spark source mass spectrometry (SSMS), glow-discharge mass spectrometry (GDMS), inductively coupled plasma (ICP) spectrometry, or other techniques should be employed.

To further assess the purity of sintered samples, a Bruker FT-IR microscope has been utilized for quantitative measurements of the transparency. The C-textured samples of said original composition have been grinded down to 0.1 mm thickness, but not optically polished. The spectra shown in the figure are the in-line transmission; the IR light propagated parallel to the C-axis, and also traveled some distance through the air, therefore additional peaks of absorption can be seen.

As was mentioned earlier, the samples sintered in air at 1500 °C were bright white (Fig. 6(c)). Some of the samples were subsequently re-sintered in the atmosphere of flowing argon at 1600 °C, and turned clear (Fig. 6(a)). Their appearance was comparable to unpolished C-plane sapphire substrates (Fig. 6(b)). Re-sintering in nitrogen did not produce a comparable effect.

The in-line infrared transmittance of the samples re-sintered in argon improved by a factor of two (Fig. 6(d)), and reached 15% in the near-infrared region. Therefore, it is possible to achieve high purity and transparency in sintered alumina despite the presence of glass-forming impurities and sintering additives in starting powders. The amount of iron oxide required for the magnetic alignment is at least an order of magnitude lower than the concentration of said impurities in commercial alumina powers.

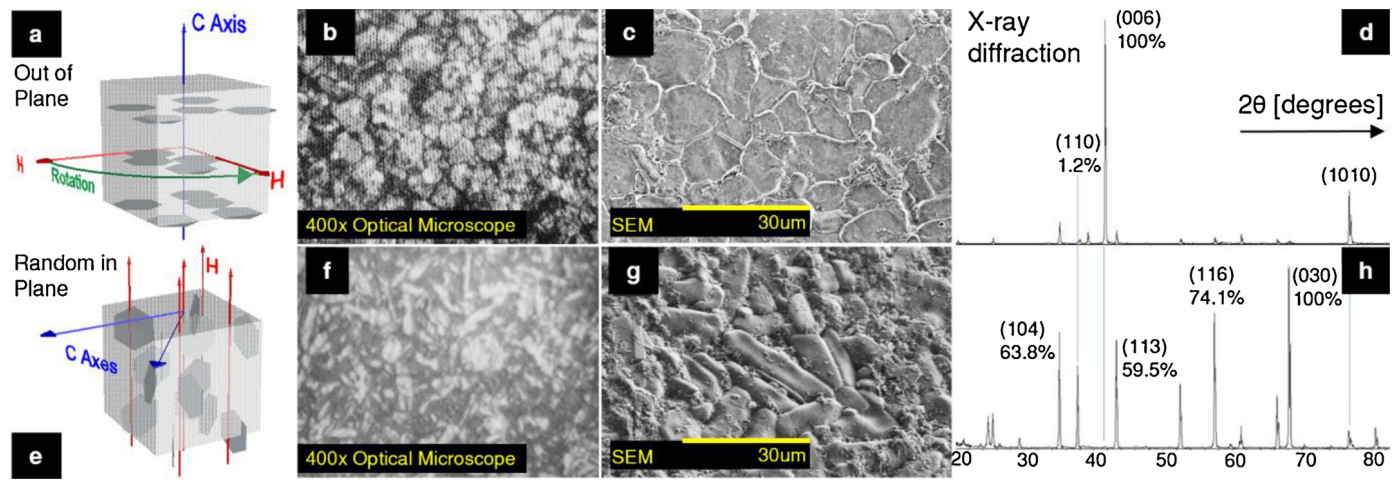

**Fig. 4.** Upper row: out of plane alignment. Lower row: random in plane alignment. (a), (e) Aligned platelets, schematic diagrams. Crystallographically aligned alumina fired at 1500C for 6 h, (b), (f) micrographs, (c), (g) SEM micrographs, (d), (h) X-ray diffraction patterns.

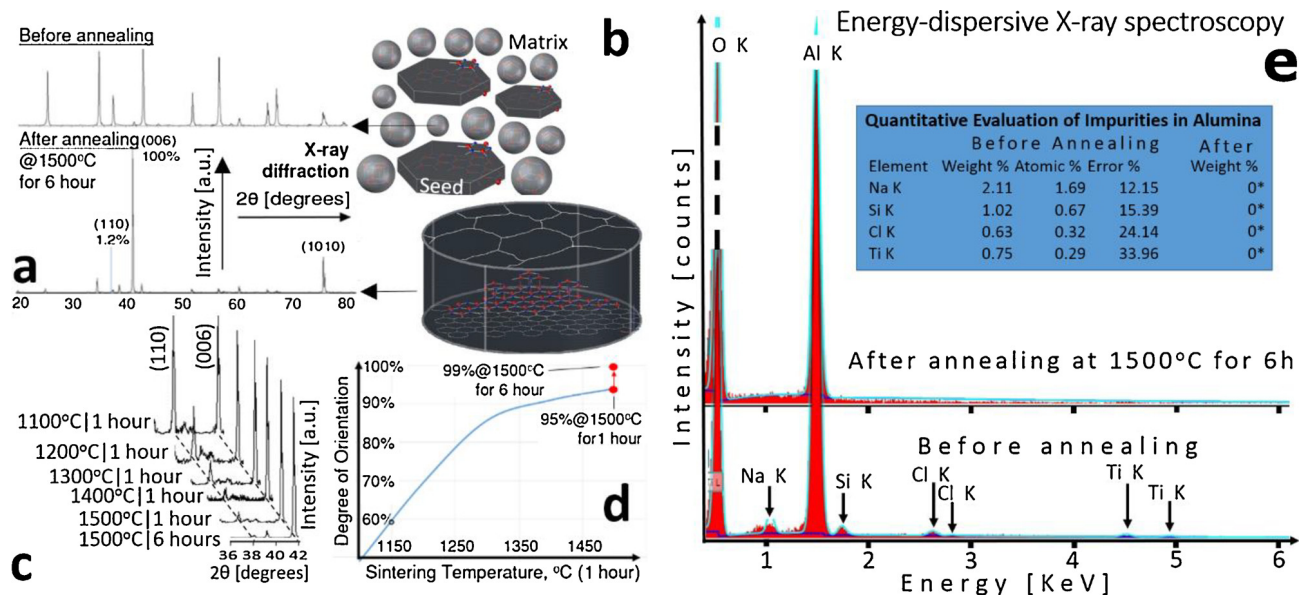

**Fig. 5.** (a) X-ray diffraction patterns of crystallographycally aligned alumina before and after annealing at 1500C for 6 h. (b) Templated gain growth (TGG), schematic diagram. (c) The evolution of X-ray diffraction patterns ((110) and (006) peaks) of crystallographycally aligned alumina as a function of sintering temperature. (d) The degree of crystallographic orientation in alumina as a function of sintering temperature. (e) Energy-dispersive X-ray spectroscopy of alumina samples before and after sintering.

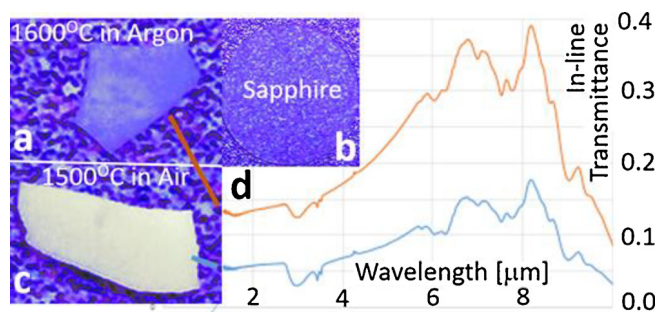

**Fig. 6.** (a) C-plane alumina sintered in flowing argon at 1600 °C. (b) C-plane sapphire substrate. (c) C-plane alumina sintered in air at 1500 °C. (d) IR in-line transmittance of 0.1 mm thick C-plane alumina samples, sintered in air (bottom) and argon (top).

The templated grain growth largely relies on a significant difference in specific surface areas between the seed and matrix. To assess the boundaries of applicability of this method, 100 nm TM-DAR alumina powder was replaced with a 1 μm powder (aluminum oxide, 99.98%, Alfa Aesar). The results discussed above were successfully repeated.

## 4. Conclusions

In summary, we present a technology that allows to precisely control the orientation of the grains in polycrystalline ceramics by application of infinitesimal magnetic fields. Ultra-high magnetic response is enabled by high aspect ratio seed crystallites coated with magnetic nanoparticles. Magnetic alignment organically complements colloidal ceramics processing and allows for the synthesis of large ceramic components with tailored crystallographic landscapes and remarkably improved functional properties. This approach relies on well-established principles, such as templated grain growth, and can be readily applied for most non-cubic engineering ceramics.

### Acknowledgement

This work was supported by the U.S. Army under Grant W911NF-15-2-0026.

### Appendix A. Supplementary data

Supplementary material related to this article can be found, in the online version, at doi:https://doi.org/10.1016/j.jeurceramsoc.2018.06.035.